\documentclass{elsart}
\usepackage{graphics}
\begin{document}
\begin{frontmatter}
\title{A Compendium of Light-Cone Quantization}
\author{Hans-Christian Pauli} 
\address{Max-Planck-Institut f\"ur Kernphysik, 
         D-69029 Heidelberg, Germany.\\ pauli@mpi-hd.mpg.de}
\date{27 August 2000/27 February 2001}  
\begin{abstract}
  For the purpose of consistent notation and easy reference the 
  most important relations in light-cone quantization are 
  compiled from a recent review \cite{BroPauPin98},
  where all further details and derivations can be found.
\end{abstract}
\hyphenation{ }
\end{frontmatter}
\begin{figure}
\resizebox{1.0\textwidth}{!}{\includegraphics{{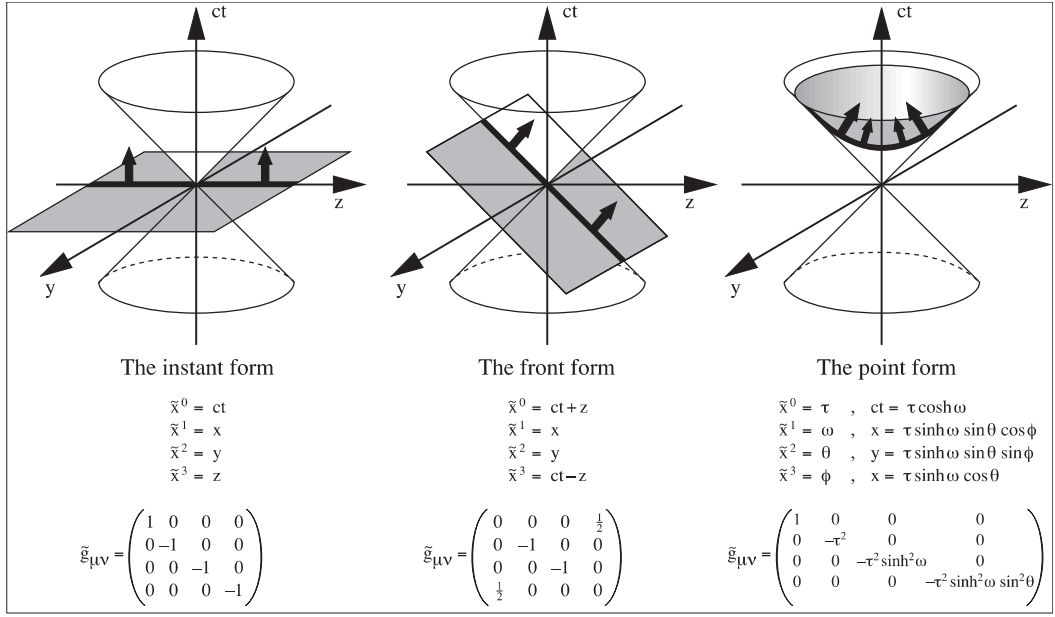}}}
\caption{\label{fig:bir2}
    Dirac's three forms of Hamiltonian dynamics.
}  \end{figure}

In the wake of the success of Feynman's action-oriented approach,
the simultaneous work by Dirac \cite{dir49} on 
the Hamiltonian approach in covariant theories 
has been forgotten for a long time.
Dirac suggests that there are essentially three forms of 
Hamiltonian dynamics: The instant, the front and the point form.
Unfortunately, one refers to the former as 
conventional `quantization' (at equal usual time) and  
`light-cone quantization' (at equal light-cone time).
The three forms have a different initialization of the
particle trajectories or the fields.
They differ by the way one parametrizes four-dimensional 
space-time as illustrated in Fig.~\ref{fig:bir2}.

%%%%%%%%%%%%%%%%%%%%%%%%%%%%%%%%%%%%%%%%%%%%%%%%%%%%%%%%%%%%%figbeg
%%%%%%%%%%%%%%%%%%%%%%%%%%%%%%%%%%%%%%%%%%%%%%%%%%%%%%%%%%%%%figend
 
\section{Four-vector conventions in the instant form }
\noindent\textbf{Lorentz vectors.} 
Contra-variant four-vectors of position $x^\mu$  
are written as
\begin {equation}
       x^\mu 
       = (x^0, x^1, x^2, x^3)
       = (t, x, y, z)
       = (x^0, \vec x_{\!\perp}, x^3)
       = (t, \vec x)
.\end {equation}
Covariant four-vectors $x_\mu$ are written as
\begin{equation}
       x_\mu 
       = (x_0, x_1, x_2, x_3)
       = (t, -x, -y, -z)
       = g_{\mu\nu} x^\nu
.\end{equation}
The metric tensors $g_{\mu\nu}$ and $g^{\mu\nu}$
lower and raise the indices, respectively, and are given 
in Fig.~\ref{fig:bir2}. 
Implicit summation over repeated 
Lorentz ($\mu,\nu,\kappa$) or  space ($i,j,k$) indices 
is understood.
Scalar products are 
\begin {equation}
       x\,p = x^\mu p_\mu 
       = x^0 p_0 + x^1 p_1+ x^2 p_2 +x^3 p_3
       = tE - \vec x\cdot \vec p
,\end{equation}
for example. The four-momentum of energy-momentum is
$ p^\mu = (p^0, p^1, p^2, p^3)$ or $p^\mu  = (E, \vec p)$. 
\\ 
\noindent\textbf{Dirac matrices.}
The $4\times4$ Dirac matrices $\gamma^\mu$ are defined by  
\begin{equation}
       \gamma^\mu\gamma^\nu +
       \gamma^\nu\gamma^\mu
       =2g^{\mu\nu}
,\end{equation}
up to unitary transformations. 
The $\gamma^k$ are anti-hermitean and $\gamma^0$ is hermitean.
Useful combinations are $\alpha^k = \gamma^0 \gamma^k$
and $\beta=\gamma^0$, as well as
\begin{equation} 
      \sigma^{\mu\nu} = 
      {i\over2}\left(
       \gamma^\mu\gamma^\nu -
       \gamma^\nu\gamma^\mu \right)
,\quad
      \gamma_5 =  \gamma^5 = 
      i \gamma^0\gamma^1\gamma^2\gamma^3
.\end{equation}
They usually are expressed in terms of the 
$2\times2$ Pauli matrices 
which are compiled here for completeness,
\begin{equation} 
      I= \left[ \begin{array} {lr}
      1 & 0 \\ 0 & 1 \end{array} \right]
,\quad
      \sigma^1= \left[ \begin{array} {lr}
      0 & 1 \\ 1 & 0 \end{array} \right]
,\quad
      \sigma^2= \left[ \begin{array} {lr}
      0 & -i \\ i & 0 \end{array} \right]
,\quad
      \sigma^3= \left[ \begin{array} {lr}
      1 & 0 \\ 0 & -1 \end{array} \right]
.\end{equation}
In \underbar{Dirac representation} 
the matrices are
\begin{eqnarray} 
      \gamma^0 &=& \left( \begin{array} {l@{\,}r}
      I & 0 \\ 0 & -I \end{array} \right)
\ , \quad
      \gamma^k = \left(\begin{array} {l@{\,}r}
      0 & \sigma^k \\ -\sigma^k & 0 \end{array} \right)
\ , \\
      \gamma_5  &=& \left( \begin{array} {l@{\,}r}
      0 & +I \\ I & 0 \end{array} \right) 
\ , \quad
     \alpha^k  = \left( \begin{array} {l@{\,}r}
      0 & \sigma^k \\ +\sigma^k& 0 \end{array} \right) 
\ , \quad
      \sigma^{ij}  = \left(\begin{array} {l@{\,}r}
      \sigma^k & 0 \\ 0 & \sigma^k \end{array} \right) 
\ . \end{eqnarray}
In \underbar{chiral representation} 
$\gamma_0$ and $\gamma_5$ are interchanged:
\begin{eqnarray}
      \gamma^0 &=& \left( \begin{array} {l@{\,}r}
      0 & +I \\ I & 0 \end{array} \right)
\ , \quad
      \gamma^k = \left( \begin{array} {l@{\,}r}
      0 & \sigma^k \\ -\sigma^k & 0 \end{array} \right)
\ , \\
      \gamma_5  &=& \left( \begin{array} {l@{\,}r}
      I & 0 \\ 0 & -I \end{array} \right) 
\ , \quad
      \alpha^k  = \left( \begin{array} {l@{\,}r}
      \sigma^k & 0 \\ 0 & -\sigma^k \end{array} \right) 
\ , \quad
      \sigma^{ij}  = \left(\begin{array} {l@{\,}r}
      \sigma^k & 0 \\ 0 & \sigma^k \end{array} \right) 
\ .\end{eqnarray}
$(i,j,k)=1,2,3$ are  used cyclically. 
%\\ 

\noindent\textbf{Projection operators.}
Combinations of Dirac matrices like 
\begin{equation}
      \Lambda_+ = {1\over2} (1+\alpha^3)
      = {\gamma^0\over2} (\gamma^0+\gamma^3)
      ,\qquad
      \Lambda_-= {1\over2} (1-\alpha^3)
      = {\gamma^0\over2} (\gamma^0-\gamma^3)
,\end{equation}
have often projector properties, particularly
\begin{equation}
      \  \quad \Lambda_++\Lambda_-={\bf 1}
      \ ,\quad \Lambda_+\Lambda_-=0
      \ ,\quad \Lambda^2_+=\Lambda_+
      \ ,\quad \Lambda^2_-=\Lambda_-
.\end{equation}
They are diagonal in the chiral  and
off-diagonal in the Dirac representation:
\begin{equation}
      (\Lambda_+)_{\rm chiral} =
      \pmatrix{1&0&0&0\cr 0&0&0&0\cr
                        0&0&0&0\cr 0&0&0&1\cr}
      \ ,\quad 
      (\Lambda_+)_{\rm Dirac} = {1\over2}
      \pmatrix{1&0&1&0\cr 0&  1&0&-1\cr
                        1&0&1&0\cr 0&-1&0&  1\cr}
.\label{eq:A14}\end{equation}
\textbf{Dirac spinors.} 
The solutions of the Dirac equations for mass $m$ particles 
are the spinors $u(p,\lambda)\equiv u_\alpha(p,\lambda)$ 
and $v(p,\lambda)\equiv v_\alpha(p,\lambda)$,
\begin{equation}
       (/\!\!\!p-m)\, u(p,\lambda)=0, \qquad
       (/\!\!\!p+m)\,v(p,\lambda)=0
.\end{equation}
They form an orthonormal set,
\begin{equation}
       \bar u(p,\lambda)u(p,\lambda')
       = - \bar v(p,\lambda')v(p,\lambda)
       = 2m\,\delta_{\lambda\lambda'} 
,\end{equation}
which is complete,
\begin{equation}
       \sum_\lambda u(p,\lambda)\bar u(p,\lambda)
       = /\!\!\!p+m 
       \ ,\ \sum_\lambda v(p,\lambda)\bar v(p,\lambda)
       = /\!\!\!p-m
.\end{equation}
Note that normalization here differs from most of the textbooks. 
The `Feynman slash' is 
$/\!\!\!p= p_\mu \gamma^\mu$.
The Gordon decomposition of currents 
\begin{equation}
       \bar u(p,\lambda)\gamma^{\mu} u(q,\lambda')
       = {1\over 2m}\bar u(p,\lambda) \Bigl( 
       (p+q)^\mu + i \sigma^{\mu\nu} (p-q)_\nu 
       \Bigr) u(q,\lambda')
\end{equation}
is often useful. The relations
\begin{equation}
      \gamma^\mu /\!\!\!a \gamma_\mu = -2a
,\qquad
      \gamma^\mu /\!\!\!a /\!\!\!b \gamma_\mu = 4ab 
,\qquad
     \gamma^\mu /\!\!\!a /\!\!\!b /\!\!\!c \gamma_\mu
      = /\!\!\!c /\!\!\!b /\!\!\!a
\end{equation}
hold identically.
With $\lambda=\pm 1$, the spin projection 
is $ s=\lambda/2$. 
\\ 
\noindent\textbf{Polarization vectors.}
The two polarization four-vectors 
$\epsilon_\mu(p,\lambda)$
are labeled by the spin projections $\lambda=\pm1$.
As solutions of the free Maxwell equations they are
orthonormal and complete: 
\begin{equation}
       \epsilon^\mu(p,\lambda)       \,
       \epsilon^\star_\mu(p,\lambda^\prime)
       = -\delta_{\lambda\lambda^\prime} 
       \ ,\qquad
       p^\mu\,\epsilon_\mu (p,\lambda)
       =0
\ .\end{equation}
The star ($^\star$) refers to complex conjugation. 
The polarization sum is 
\begin{equation}
       d_{\mu\nu}(p)=\sum\limits_{\lambda}
       \epsilon_\mu (p,\lambda)
       \epsilon^\star_\nu (p,\lambda)
       = - g_{\mu\nu}
       +{\eta_\mu p_\nu+\eta_\nu p_\mu
       \over p^\kappa \eta_\kappa}
\ ,\end{equation}
with the null vector $\eta^\mu\eta_\mu = 0$ given below.

\section{Additional front-form conventions (LB)\cite{leb80}}
\label{sec:uv}
\noindent\textbf{Lorentz vectors.} 
Contra-variant four-vectors of position $x^\mu$
are written as 
\begin {equation}
       x^\mu 
       = (x^+, x^1, x^2, x^-)
       = (x^+, \vec x_{\!\perp}, x^-)
\ . \end {equation}
Its time-like and space-like components are related to
the instant form by 
\begin {equation}
       x^+ = x^0 + x^3
\quad{\rm and}\quad
       x^- = x^0 - x^3
\ , \end {equation}
respectively, and referred to as `light-cone time'
and `light-cone space'. The covariant vectors are
obtained by $x_\mu=g_{\mu\nu}x^\nu$, 
with the metric tensor(s) 
\begin {equation}
       g^{\mu\nu}=\pmatrix{0&0&0&2 \cr 
       0&-1&0&0 \cr 0&0&-1&0 \cr 2&0&0&0 \cr }
       \qquad{\rm and}\qquad
       g_{\mu\nu}=\pmatrix{0&0&0&\frac{1}{2} \cr 
       0&-1&0&0 \cr 0&0&-1&0 \cr \frac{1}{2}&0&0&0 \cr }
\ .\end {equation}
Scalar products $x p = x^\mu p_\mu$ are 
\begin {equation}
       x^\mu p_\mu 
       = x^+ p_+ + x^- p_- + x^1 p_1+ x^2 p_2 
       =  {1\over2}(x^+ p^- + x^- p^+) 
       - \vec x_{\!\perp} \vec p_{\!\perp}
\ .\end{equation}
All other four-vectors including $\gamma^\mu$ 
are treated correspondingly.
\\
\noindent\textbf{Dirac matrices.}
In Dirac representation of the $\gamma$-matrices holds
\begin {equation}
       \gamma^+\gamma^+ =
       \gamma^- \gamma^-  = 0
\ .\end {equation}
Alternating products are for example
\begin{equation}
       \gamma^+\gamma^-\gamma^+ = 4\gamma^+
       ,\qquad 
       \gamma^-\gamma^+\gamma^-  = 4\gamma^- 
\ .\end{equation}
\textbf{Projection operators.}
The projection matrices become
\begin{equation}
       \Lambda_+={1\over2}\gamma^0\gamma^+ 
       ={1\over4}\gamma^-\gamma^+ 
       ,\qquad 
       \Lambda_- ={1\over2}\gamma^0\gamma^- 
       ={1\over4}\gamma^+\gamma^- 
.\end{equation}
%%%%%%%%%%
\begin{table} [t]
\caption [uudir] {\label {tab:uudir}
   {Matrix elements of Dirac spinors} 
   $\bar u(p){\mathcal M}u(q)$. 
   \par\vskip 1em}
\begin{center}
\begin{tabular}{|c@{}|c|c|} 
\hline 
    $\displaystyle \rule[2ex]{0ex}{2ex} {\mathcal  M} \rule[-3ex]{0ex}{2ex} $ 
  & $\displaystyle \frac{\bar u(p){\mathcal  M}u(q)} 
    {\sqrt{p^+q^+}} \,\delta_{\lambda_p,\lambda_q}$
  & $\displaystyle \frac{\bar u(p){\mathcal  M}u(q)}
    {\sqrt{p^+q^+}}\,\delta_{\lambda_p,-\lambda_q} $
\\ \hline \hline
   $\displaystyle \rule[1ex]{0ex}{2ex}\gamma^+\rule[-0.0ex]{0ex}{2ex} $    
  &$2$   &$0$
\\ 
   $\displaystyle \rule[2ex]{0ex}{2ex} \gamma^- \rule[-2.5ex]{0ex}{2ex} $ 
  &$\displaystyle {2\over p^+ q^+}
     \left(\vec p_{\!\perp} \!\cdot\!\vec q_{\!\perp} + m^2 
     +i\lambda_q \vec p_{\!\perp}\!\wedge\!\vec q_{\!\perp}\right)$ 
  &$\displaystyle {2m\over p^+ q^+} \left( 
     p_{\!\perp}(\lambda_q) - q_{\!\perp}(\lambda_q) 
     \right) $
\\ 
   $\displaystyle \rule[2ex]{0ex}{2ex} 
    \vec\gamma_{\!\perp}\!\cdot\!\vec a_{\!\perp} 
    \rule[-2.5ex]{0ex}{2ex} $ 
  &$\displaystyle \vec a_{\!\perp}\!\cdot\!\left(
    {\vec p_{\!\perp}\over p^+}+{\vec q_{\!\perp}\over q^+}
    \right) - i\lambda_q 
    \vec a_{\!\perp}\!\wedge\!\left(
    {\vec p_{\!\perp}\over p^+}-{\vec q_{\!\perp}\over q^+}
    \right) $ 
  &$\displaystyle - a_{\!\perp}(\lambda_q)\left(  
     {m\over p^+} - {m\over q^+} \right) $
\\ \hline
   $\displaystyle \rule[2ex]{0ex}{2ex} 1\rule[-2.5ex]{0ex}{2ex} $ 
  &$\displaystyle {m\over p^+}+{m\over q^+}$
  &$\displaystyle 
       {p_{\!\perp}(\lambda_q)\over q^+} 
     - {q_{\!\perp}(\lambda_q)\over p^+} $
\\ \hline 
   $\displaystyle \rule[1.5ex]{0ex}{2ex} 
    \gamma^-\,\gamma^+\,\gamma^- \rule[-2.5ex]{0ex}{2ex} $ 
  &$\displaystyle {8\over p^+ q^+}
     \left(\vec p_{\!\perp} \!\cdot\!\vec q_{\!\perp} + m^2 
     +i\lambda_q \vec p_{\!\perp}\!\wedge\!\vec q_{\!\perp}\right)$
  &$\displaystyle {8m\over p^+ q^+} \left( 
     p_{\!\perp}(\lambda_q) - q_{\!\perp}(\lambda_q) 
     \right) $
\\ 
   $\displaystyle \rule[1.0ex]{0ex}{2ex} \gamma^-\,\gamma^+\,
     \vec\gamma_{\!\perp}\!\cdot\!\vec a_{\!\perp}
     \rule[-2.0ex]{0ex}{2ex}$
  &$\displaystyle {4\over p^+} 
     \left(\vec a_{\!\perp}\!\cdot\!\vec p_{\!\perp}- i\lambda_q 
     \vec a_{\!\perp}\!\wedge\!\vec p_{\!\perp}\right) $
  &$\displaystyle - {4m\over p^+}a_{\!\perp}(\lambda_q)$
\\ 
   $\displaystyle \rule[2ex]{0ex}{2ex} 
    \vec a_{\!\perp}\!\cdot\!\vec\gamma_{\!\perp}
     \,\gamma^+\,\gamma^-\rule[-2.5ex]{0ex}{2ex} $
  &$\displaystyle {4\over q^+} 
     \left(\vec a_{\!\perp}\!\cdot\!\vec q_{\!\perp} + i\lambda_q 
     \vec a_{\!\perp}\!\wedge\!\vec q_{\!\perp}\right) $
  &$\displaystyle \phantom{-}{4m\over q^+}
    a_{\!\perp}(\lambda_q)$
\\ 
   $\displaystyle \vec a_{\!\perp}\!\cdot\!\vec\gamma_{\!\perp}
   \,\gamma^+\,\vec\gamma_{\!\perp}\!\cdot\!\vec b_{\!\perp}$
  &$\displaystyle 2\Big(\vec a_{\!\perp}\!\cdot\!\vec b_{\!\perp} 
   + i\lambda_q\vec a_{\!\perp}\!\wedge\!\vec b_{\!\perp}\Big)$
  &$\displaystyle \rule[1.0ex]{0ex}{2ex} 0 \rule[-1.5ex]{0ex}{2ex} $
\\  \hline \hline
   \multicolumn{3}{|l|} { {\rm Notation:\ }\qquad
   $\displaystyle \rule[1.0ex]{0ex}{2ex} \lambda = \pm 1$,\quad
   $\displaystyle a_{\!\perp}(\lambda) 
     = -\lambda a_ x - i a_y $\hfill}
\\   \multicolumn{3}{|l|} { \phantom{\rm Notation:\ }\qquad
      $\displaystyle \vec a_{\!\perp}\!\cdot\!\vec b_{\!\perp} 
     = a_x b_ x + a_y b_y $, \quad
   $\displaystyle \vec a_{\!\perp}\!\wedge\!\vec b_{\!\perp} 
     = a_x b_y - a_y b_ x $.\hfill} 
\\ \multicolumn{3}{|l|} {{\rm Symmetries:\ } \qquad
   $\displaystyle \bar v(p) \,v(q) = - \bar u(q) \,u(p)$, \quad
   $\displaystyle \bar v(p) \,\gamma^\mu\,v(q) 
   =   \bar u(q) \,\gamma^\mu\,u(p)$, \hfill }
\\ \multicolumn{3}{|l|} {\phantom{{\rm Symmetries:\ }\qquad}
   \rule[-1.5ex]{0ex}{2ex} 
   $\displaystyle \bar v(p) 
   \,\gamma^\mu\gamma^\nu\gamma^\rho\,v(q) 
   =   \bar u(q) 
   \,\gamma^\rho\gamma^\nu\gamma^\mu\,u(p)$.\hfill }
\\ \hline 
\end{tabular}
\end{center}
\end{table}
%%%%%%%%%%%
\textbf{Dirac spinors.} 
The Lepage-Brodsky representation is particularly simple: 
\begin{eqnarray}
       u(p, \lambda) = {1\over \sqrt{p^+} } 
       \left(p^+ + \beta m + \vec \alpha_{\!\perp} 
       \vec p_{\!\perp}\right) \times
       \cases{  \chi (\uparrow), &for $\lambda=+1$, \cr
       \chi (\downarrow), &for $\lambda=-1$, \cr} 
\\
       v(p, \lambda) = {1\over \sqrt{p^+} } 
       \left(p^+ - \beta m + \vec \alpha_{\!\perp} 
       \vec p_{\!\perp}\right) \times
       \cases{ \chi (\downarrow), &for $\lambda=+1$, \cr
       \chi (\uparrow), &for $\lambda=-1$. \cr}
\end{eqnarray}
The two $ \chi $-spinors are
\begin{equation}
      \chi (\uparrow) = {1\over\sqrt{2} } \, 
      \left(\begin{array} {r}
       1 \\  0 \\ 1 \\ 0 \end{array}\right)
      \qquad{\rm and}\qquad
      \chi (\downarrow) = {1\over\sqrt{2}} \, 
      \left(\begin{array} {r}
       0 \\  1 \\ 0 \\ -1 \end{array}\right)
.\label{eq:B10}\end{equation}
%\\
\noindent\textbf{Tables of Dirac spinors.}
Matrix elements of Dirac spinors are given in 
Tables~\ref{tab:uudir} and \ref{tab:uvdir}.
As a particular application the Lorentz invariant spinor factor 
\begin{equation}
   S = j ^\mu \overline j _\mu =
   \left[ \overline u (k_{1},\lambda_{1})\gamma^\mu
   u(k_{1}^\prime,\lambda_{1}^\prime)\right] \, 
   \left[ \overline v(k_{2}^\prime,\lambda_{2}^\prime) \gamma_\mu 
    v(k_{2},\lambda_{2})\right] 
\end{equation}
is calculated explicitly. 
It appears for example in the $q\bar q$-scattering amplitude.
In helicity space it can be understood as the matrix
$\langle\lambda_1 ,\lambda_2\vert S\vert\lambda_1',\lambda_2'\rangle$ 
whose matrix elements are functions of $x,\vec k _{\!\perp}$
and $x',\vec k _{\!\perp}'$. 
For convenience, $S$ is calculated here as $S=2T\sqrt{x(1-x)x'(1-x')}$, 
{\it i.e.}
\begin{equation}
  \langle
  \lambda_{q},\lambda_{\bar q}\vert T \vert 
  \lambda_{q}^\prime,\lambda_{\bar q}^\prime\rangle =
  \frac {1}{2}
  \frac{\left[\overline u (k_1,\lambda_1)\gamma^\mu
   u(k_1^\prime,\lambda_1^\prime)\right]}
   {\sqrt{xx'} } 
  \frac{\left[\overline v (k_2^\prime,\lambda_2^\prime)
  \gamma_\mu v (k_2,\lambda_2) \right]}
   {\sqrt{(1-x)(1-x')} } 
.\end{equation}
It is often useful to arrange
$S$ or $T$ as a matrix in helicity space,
\begin{equation}
  \langle
  \lambda_{q},\lambda_{\bar q}\vert T \vert 
  \lambda_{q}^\prime,\lambda_{\bar q}^\prime\rangle =
  \bordermatrix{  &\uparrow\downarrow&\downarrow\uparrow 
                  &\uparrow\uparrow  &\downarrow\downarrow   \cr 
  \uparrow\downarrow     & T_{11} & T_{12} & T_{13} & T_{14} \cr 
  \downarrow\uparrow     & T_{21} & T_{22} & T_{23} & T_{24} \cr 
  \uparrow\uparrow       & T_{31} & T_{32} & T_{33} & T_{34} \cr 
  \downarrow\downarrow   & T_{41} & T_{42} & T_{43} & T_{44} \cr }
.\end{equation}
With $y\equiv 1-x$ the diagonal elements are
\begin{eqnarray}
    T_{11} &=& 
    \frac{m^2_1}{xx'} + \frac{m^2_2}{yy'} + 
    \frac{\vec{k }^2_{\!\perp}}{x y} + 
    \frac{\vec{k'}^2_{\!\perp}}{x'y'}     
    + 
    \frac{\vec k _{\!\perp}\cdot\vec k'_{\!\perp} + i
    \vec k _{\!\perp}\!\!\wedge\!\vec k'_{\!\perp} }{xx'}
    + 
    \frac{\vec k _{\!\perp}\cdot\vec k'_{\!\perp} - i
    \vec k _{\!\perp}\!\!\wedge\!\vec k'_{\!\perp} }{yy'}     
,\\
    T_{22} &=& T_{11} 
,\\
    T_{33} &=& 
    \frac{m^2_1}{xx'}+\frac{m^2_2}{yy'}  +
    \frac {\vec{k}_{\!\perp}\cdot\vec k'_{\!\perp} +i
           \vec{k}_{\!\perp}\!\!\wedge\!\vec k'_{\!\perp}}{xyx'y'}
,\\
    T_{44} &=& T_{33} 
.\end{eqnarray}
The off-diagonal matrix elements become
\begin{eqnarray}
    T_{12} &=& - m_1m_2\frac{(x - x')^2}{xy x'y'} 
    ,\hskip 4.6em
    T_{21}  =  - m_1m_2\frac{(x - x')^2}{xy x'y'} 
,\\
    T_{13} &=& \frac{m_2}{yy'}\ x \left[
    \frac{k _{\!\perp}(\uparrow)}{x} - 
    \frac{k'_{\!\perp}(\uparrow)}{x'}\right]
    ,\qquad 
    T_{31}  =  \frac{m_2}{yy'}\ x'\left[
    \frac{k _{\!\perp}(\downarrow)}{x} - 
    \frac{k'_{\!\perp}(\downarrow)}{x'}\right]
,\\
    T_{14} &=& \frac{m_1}{xx'}\ y \left[
    \frac{k _{\!\perp}(\downarrow)}{y} - 
    \frac{k'_{\!\perp}(\downarrow)}{y'}\right]
    ,\qquad 
    T_{41}  =  \frac{m_1}{xx'}\ y'\left[
    \frac{k _{\!\perp}(\uparrow)}{y} - 
    \frac{k'_{\!\perp}(\uparrow)}{y'}\right]
,\\
    T_{23} &=& \frac{m_1}{xx'}\ y \left[
    \frac{k _{\!\perp}(\uparrow)}{y} - 
    \frac{k'_{\!\perp}(\uparrow)}{y'}\right]
    ,\qquad 
    T_{32}  =  \frac{m_1}{xx'}\ y'\left[
    \frac{k _{\!\perp}(\downarrow)}{y} - 
    \frac{k'_{\!\perp}(\downarrow)}{y'}\right]
,\\
    T_{24} &=& \frac{m_2}{yy'}\ x \left[
    \frac{k _{\!\perp}(\downarrow)}{x} - 
    \frac{k'_{\!\perp}(\downarrow)}{x'}\right]
    ,\qquad 
    T_{42}  =  \frac{m_2}{yy'}\ x'\left[
    \frac{k _{\!\perp}(\uparrow)}{x} - 
    \frac{k'_{\!\perp}(\uparrow)}{x'}\right]
,\\
    T_{34} &=& 0,\hskip 11.5em T_{43}  =  0
,\end{eqnarray}
where 
$k _{\!\perp}(\uparrow)   = -k _{\!\perp x} -ik _{\!\perp y}$ and
$k _{\!\perp}(\downarrow) =  k _{\!\perp x} -ik _{\!\perp y}$. 
\\
Close to kinematic equilibrium, {\it i.e.} for
$x\sim x'= \overline x$ and
$\vec k _{\!\perp}\sim \vec k _{\!\perp}'=0$,
the particles are essentially at rest, and $S$
is diagonal ($\overline x=m_1/(m_1+m_2)$):
\begin{equation}
   \langle\lambda_1,\lambda_2\vert S\vert\lambda_1',\lambda_2'\rangle =
   4 m_1 m_2\,
   \delta_{\lambda_1,\lambda_1'}\,\delta_{\lambda_2,\lambda_2'}
.\end{equation}
Far-off equilibrium $S$ is also diagonal; 
all matrix elements vanish essentially 
as compared to the two leading ones 
(for $x\sim x'$ and $k'_{\!\perp}\gg k_{\!\perp} \gg m$):
\begin{equation}
   \langle\uparrow\downarrow\vert S\vert\uparrow\downarrow\rangle =
   \langle\downarrow\uparrow\vert S\vert\downarrow\uparrow\rangle = 
   2\vec{k'}^2_{\!\perp}
.\end{equation}
%
%%%%%%%%%%%%%%%
\begin{table} [t]
\caption [uvdir] {\label {tab:uvdir}
   {Matrix elements of Dirac spinors}
   $\bar v(p){\mathcal M}u(q)$. 
   \par\vskip 1em}
\begin{center}
\begin{tabular}{|c@{}|c|c|} 
\hline 
    $\displaystyle \rule[2ex]{0ex}{2ex} {\mathcal  M} \rule[-3ex]{0ex}{2ex} $ 
  & $\displaystyle \frac{\bar v(p){\mathcal  M}u(q)} 
    {\sqrt{p^+q^+}} \,\delta_{\lambda_p,\lambda_q}$
  & $\displaystyle \frac{\bar v(p){\mathcal  M}u(q)}
    {\sqrt{p^+q^+}}\,\delta_{\lambda_p,-\lambda_q} $
\\ \hline \hline
   $\displaystyle \rule[1ex]{0ex}{2ex}\gamma^+\rule[-0.0ex]{0ex}{2ex} $    
   &$0$ &$2$   
\\ 
   $\displaystyle \rule[2ex]{0ex}{2ex} \gamma^- \rule[-2.5ex]{0ex}{2ex} $ 
  &$\displaystyle {2m\over p^+ q^+} 
    \left( p_{\!\perp} (\lambda_q) +
           q_{\!\perp} (\lambda_q) \right) $
  &$\displaystyle  {2\over p^+ q^+}
     \left(\vec p_{\!\perp} \!\cdot\!\vec q_{\!\perp} - m^2 
     +i\lambda_q \vec p_{\!\perp}\!\wedge\!\vec q_{\!\perp}\right)$ 
\\ 
   $\displaystyle \rule[2ex]{0ex}{2ex} 
    \vec\gamma_{\!\perp}\!\cdot\!\vec a_{\!\perp} 
    \rule[-2.5ex]{0ex}{2ex} $ 
  &$\displaystyle a_{\!\perp}(\lambda_q)\left(
     {m\over p^+} + {m\over q^+} \right) $
  &$\displaystyle \vec a_{\!\perp}\!\cdot\!\left(
    {\vec p_{\!\perp}\over p^+}+{\vec q_{\!\perp}\over q^+}
    \right) - i\lambda_q 
    \vec a_{\!\perp}\!\wedge\!\left(
    {\vec p_{\!\perp}\over p^+}-{\vec q_{\!\perp}\over q^+}
    \right) $ 
\\ \hline
   $\displaystyle \rule[2ex]{0ex}{2ex} 1\rule[-2.5ex]{0ex}{2ex} $ 
  &$\displaystyle 
     {p_{\!\perp}(\lambda_q)\over p^+} + 
     {q_{\!\perp}(\lambda_q)\over q^+}  $
  &$\displaystyle -{m\over p^+}+{m\over q^+}$
\\ \hline 
   $\displaystyle \rule[1.5ex]{0ex}{2ex} 
    \gamma^-\,\gamma^+\,\gamma^- \rule[-2.5ex]{0ex}{2ex} $ 
  &$\displaystyle {8m\over p^+ q^+} 
    \left( p_{\!\perp}(\lambda_q) 
         + q_{\!\perp}(\lambda_q) \right) $
  &$\displaystyle {8\over p^+ q^+}
     \left(\vec p_{\!\perp} \!\cdot\!\vec q_{\!\perp} - m^2 
     +i\lambda_q \vec p_{\!\perp}\!\wedge\!\vec q_{\!\perp}\right)$
\\ 
   $\displaystyle \rule[1.0ex]{0ex}{2ex} \gamma^-\,\gamma^+\,
     \vec\gamma_{\!\perp}\!\cdot\!\vec a_{\!\perp}
     \rule[-2.0ex]{0ex}{2ex}$
  &$\displaystyle {4m\over p^+} a_{\!\perp}(\lambda_q)$
  &$\displaystyle {4\over p^+} 
     \left(\vec a_{\!\perp}\!\cdot\!\vec p_{\!\perp}- i\lambda_q  
     \vec a_{\!\perp}\!\wedge\!\vec p_{\!\perp}\right) $
\\ 
   $\displaystyle \rule[2ex]{0ex}{2ex} 
    \vec a_{\!\perp}\!\cdot\!\vec\gamma_{\!\perp}
     \,\gamma^+\,\gamma^-\rule[-2.5ex]{0ex}{2ex} $
  &$\displaystyle {4m\over q^+}
    a_{\!\perp}(\lambda_q)$
  &$\displaystyle {4\over q^+} 
     \left(\vec a_{\!\perp}\!\cdot\!\vec q_{\!\perp} + i\lambda_q 
     \vec a_{\!\perp}\!\wedge\!\vec q_{\!\perp}\right) $
\\ 
   $\displaystyle \vec a_{\!\perp}\!\cdot\!\vec\gamma_{\!\perp}
   \,\gamma^+\,\vec\gamma_{\!\perp}\!\cdot\!\vec b_{\!\perp}$
  &$\displaystyle \rule[1.0ex]{0ex}{2ex} 0 \rule[-1.5ex]{0ex}{2ex} $
  &$\displaystyle 2\Big(\vec a_{\!\perp}\!\cdot\!\vec b_{\!\perp} 
   + i\lambda_q\vec a_{\!\perp}\!\wedge\!\vec b_{\!\perp}\Big)$
\\  \hline \hline
   \multicolumn{3}{|l|} { {\rm Notation:\ }\qquad
   $\displaystyle \rule[1.0ex]{0ex}{2ex} \lambda = \pm 1$,
   \quad   $\displaystyle a_{\!\perp}(\lambda) 
     = -\lambda a_ x - i a_y $ \hfill}
\\   \multicolumn{3}{|l|} { \phantom{\rm Notation:\ }\qquad
      $\displaystyle \vec a_{\!\perp}\!\cdot\!\vec b_{\!\perp} 
     = a_x b_ x + a_y b_y $, \quad
   $\displaystyle \vec a_{\!\perp}\!\wedge\!\vec b_{\!\perp} 
     = a_x b_y - a_y b_ x $.\hfill} 
\\ \multicolumn{3}{|l|} {{\rm Symmetries:\ } \qquad
   $\displaystyle \bar v(p) \,v(q) 
   = - \bar u(q) \,u(p)$, \quad
   $\displaystyle \bar v(p) \,\gamma^\mu\,v(q) 
   =   \bar u(q) \,\gamma^\mu\,u(p)$, \hfill }
\\ \multicolumn{3}{|l|} {\phantom{{\rm Symmetries:\ }\qquad}
   \rule[-1.5ex]{0ex}{2ex} 
   $\displaystyle \bar v(p) 
   \,\gamma^\mu\gamma^\nu\gamma^\rho\,v(q) 
   =   \bar u(q) 
   \,\gamma^\rho\gamma^\nu\gamma^\mu\,u(p)$.\hfill }
\\ \hline 
\end{tabular}
\end{center}
\end{table}
%%%%%%%
\textbf{Polarization vectors.}
The null vector is 
\begin{equation}
       \eta^\mu =
       \left( 0, \vec 0, 2\right)
.\end{equation}
The transversal polarization vectors 
$\vec\epsilon_{\!\perp} (\uparrow) = -\frac{1}{\sqrt{2}}\,(1,i)$ and 
$\vec\epsilon_{\!\perp}(\downarrow) = \frac{1}{\sqrt{2}}\,(1,-i)$ 
are the same as in Bj\o rken-Drell convention:
Circular polarization has the spin projections 
$\lambda=\pm1= \uparrow\downarrow$. 
With the unit vectors $\vec e_x$ and $\vec e_y$
in $p_x$- and $p_y$-direction, respectively,
one writes collectively
\begin{equation}
       \vec \epsilon_{\!\perp} (\lambda)   =
      {-1\over\sqrt{2}}(\lambda\vec e_x + i \vec e_y)
.\end{equation}
The light-cone gauge $A^+=0$ induces 
$\epsilon^+ (p, \lambda) = 0$, thus  
\begin{equation}
       \epsilon^\mu (p, \lambda) =
       \left( 0, \vec \epsilon_{\!\perp}(\lambda),  
       {2\vec \epsilon_{\!\perp} (\lambda)
       \vec p_{\!\perp}\over p^+} \right)
,\end{equation}
which satisfies the transversality condition 
$ p_\mu\epsilon^\mu (p, \lambda)=0$ identically. 
\section{Canonical field theory for Quantum Chromodynamics}

If one replaces each local gauge field 
$ A^\mu (x)$ in QED by the {matrix} ${\bf A} ^\mu (x) $, 
\begin {equation}
   A ^\mu \longrightarrow
  ({\bf A}^\mu)_{cc^\prime} = {1\over2} \pmatrix{
   {1\over\sqrt3} A^\mu_8+ A^\mu_3  
  & A^\mu_1-i A^\mu_2    
  & A^\mu_4-i A^\mu_5 \cr
    A^\mu_1+i A^\mu_2    
  &{1\over\sqrt3} A^\mu_8- A^\mu_3  
  & A^\mu_6-i A^\mu_7 \cr
    A^\mu_4+i A^\mu_5    
  & A^\mu_6+i A^\mu_7      
  &-{2\over\sqrt3} A^\mu_8 \cr} 
,\label{eq:2.25} \end {equation}
one generates  
{Quantum Chromodynamics} with its eight real valued 
{color vector potentials} $ A ^\mu _a$, enumerated by the
{glue index} $a = 1,2, \dots, 8$. 
These matrices are all {hermitean} and 
{traceless} since the trace can always be absorbed into 
an Abelian U(1) gauge theory. 
Unitary transformations of this matrix belong to the class of
{special unitary $3\times 3$ matrices}, SU(3).
In order to make sense of expressions like 
$ \overline \Psi {\bf A} ^\mu \Psi $ the {quark fields} $\Psi (x)$ 
must carry a {color index} $c = 1,2,3$ which is usually 
suppressed in the color triplet spinor $\Psi _{c, \alpha } (x)$.

More generally for SU(N), the vector potentials $ {\bf A}^\mu $
are hermitian and traceless $N \times N$ matrices.
All such matrices can be parametrized  
$ {\bf A} ^\mu \equiv T ^a _{cc ^\prime} A ^\mu _a$.
The color index $c$ (or $c^\prime$) runs now from 1 to $n_c$, 
and correspondingly the gluon index $a$ (or $r,s,t$) 
from 1 to $n_c^2-1$. Both are implicitly summed over, 
with no distinction of lowering or raising them.
The color matrices $ T ^a_{cc ^\prime}$ obey 
\begin {equation}
  \Bigl[ T ^r, T ^s \Bigr] _{cc ^\prime} 
  = i f ^{rsa}  T _{cc ^\prime} ^a 
  \qquad\quad {\rm and } \qquad
  {\rm Tr}\ \bigl(T ^r  T ^s \bigr) = \frac{1}{2}\ \delta _r ^s 
.\label{eq:2.26}\end   {equation}
The {\it structure constants} $\displaystyle f ^{rst}$ for SU(3)
are tabulated in the literature. 
For SU(2) they are the totally antisymmetric tensor
$\displaystyle \epsilon _{rst}$, since 
$T ^a = {1\over 2} \sigma ^a $ with $\sigma ^a$ being the Pauli 
matrices. For SU(3), the $ T ^a $ are related to the Gell-Mann 
matrices $\lambda ^a$ by $ T ^a = {1\over 2} \lambda ^a$, 
see also Eq.(\ref{eq:2.25}).
The gauge-invariant Lagrangian density for QCD or SU(N) is then
\begin{equation}
    {\mathcal L} = - \frac{1}{2} 
    {\rm Tr} \bigl({\bf F}^{\mu\nu} {\bf F}_{\mu \nu} \bigr)
 + \frac{1}{2}  \bigl[ \overline \Psi \bigl(i\gamma^\mu {\bf D} _\mu 
   - {\bf m} \bigr)  \Psi   + \ {\rm h.c.} \bigr] 
.\end{equation}
The factor $\frac{1}{2}$ is because of 
the trace convention in Eq.(\ref{eq:2.26}). 
The mass matrix ${\bf m} = m \delta_{cc^\prime}$ is diagonal
in color space. The matrix notation is particularly
suited for establishing gauge invariance,
with the unitary operators $ {\bf U}$ 
now being $N \times N $ matrices, 
hence {\it non-Abelian} gauge theory.
Gauge invariance generates an extra term in
the {color-electro-magnetic} fields
\begin {equation}
   {\bf F}^{\mu\nu} \equiv \partial^\mu  {\bf A}^\nu  
   - \partial^\nu  {\bf A}^\mu 
   + i g \bigl[ {\bf A}^\mu, {\bf A}^\nu\bigr] 
.\label{eq:2.31}\end{equation}
The covariant derivative {\it matrix} finally is
$ {\bf D} ^\mu _{cc^\prime} =
  \delta_{cc^\prime} \partial^\mu 
  + i g {\bf A}^\mu _{cc^\prime} $,
and the canonical energy-momentum stress tensor becomes
\begin {equation}
    T ^{\mu\nu} = 2{\rm Tr} \bigl({\bf F}^{\mu\kappa} 
   {\bf F}_\kappa^{\phantom{\kappa}\nu}\bigr)
  + {1\over2} \bigl[ i \overline \Psi \gamma^\mu  {\bf D} ^\nu \Psi  
  + \ {\rm h.c.} \bigr]  - g^{\mu\nu} {\mathcal L}  
.\end {equation} 
Integrating  it over a space-like hyper-surface, with the 
surface elements $ d \omega _\lambda $ and the 
(finite) volume $\Omega$ defined most conveniently 
in terms of the totally antisymmetric tensor 
$ \epsilon _{\lambda \mu \nu \rho}$ ($ \epsilon _{+12-} = 1$),
\begin{equation} 
  d \omega _\lambda = {1 \over 3 !} 
  \epsilon _{\lambda \mu \nu \rho} d x ^\mu d x ^\nu d x ^\rho
  \qquad {\rm and }\quad
  \Omega = \int d \omega_+ = \int d x ^1 d x ^2 d x ^-
,\end{equation} 
respectively, one arrives at the canonical energy-momentum four-vector 
\begin{equation} 
    P^\nu = \int_\Omega \! d\omega_+ 
    \ \biggl( 2 {\rm Tr} ({\bf F}^{+\kappa} 
    {\bf F}_\kappa^{\phantom{\kappa}\nu})
  - g^{+\nu} {\mathcal L}  + {1\over2} 
  \Bigl[ i \overline \Psi \gamma^+ {\bf D} ^\nu \Psi  
  + \ {\rm h.c.} \Bigr] \biggr) 
.\label{eq:2.39}\end{equation}
Note that both $T ^{\mu\nu}$ and $P^\nu$ 
are {\it manifestly gauge-invariant}, and that
all of this holds for SU(N), in fact that it holds 
for $d+1$ dimensions.

Since $P^\nu$ is gauge-independent and a constant of motion,
it can be evaluated in the light-cone gauge $A^+=0$
and at the fixed light-cone time $x^+=0$.
As initial condition one uses the free fields.
For the fermions they are
\begin{equation}
   \widetilde\Psi  _{\alpha cf} (x) =
   \sum _\lambda\!\int\!\! {
   dk^+ d^2k_{\!\bot}\over \sqrt{2k^+(2\pi)^3}} \left( 
   b_q         u_\alpha(k,\lambda) e^{-ikx} +
   d_q^\dagger v_\alpha(k,\lambda) e^{+ikx} \right)
.\end{equation}
Each fermion is specified by the five numbers 
$q = \left( k^+,{k_{\!\perp}}_x,{k_{\!\perp}}_x,\lambda,c,f\right)$,
{\it i.e.} by the three momenta,the helicity, color, and flavor,
repectively. 
The fermion fields $\widetilde\Psi$ become operator valued by
the anti-commutation relations 
\begin {equation} 
   \left\{b_q, b^\dagger_{q'}\right\} = 
   \left\{d_q, d^\dagger_{q'}\right\} = 
   \delta (k^+-k^{+\,\prime}) 
   \delta ^{(2)}(\vec k_{\!\bot}-\vec k_{\!\bot} ^{\,\prime}) 
   \delta _\lambda ^{\lambda ^\prime} 
   \delta _c^{c^\prime} 
   \delta _f^{f^\prime} 
.\end {equation}
For the gauge bosons, the free fields are
\begin{equation}
   \widetilde A _\mu^a (x) =
   \sum _\lambda\!\int\!\!{
   dk^+ d^2k_{\!\bot}\over \sqrt{2k^+(2\pi)^3}} \left(
   a _q         \epsilon_\mu      (k,\lambda) e^{-ikx} +
   a _q^\dagger \epsilon_\mu^\star(k,\lambda) e^{+ikx} \right) 
.\end{equation}
Each boson is specified by the four numbers 
$q = \left( k^+,{k_{\!\perp}}_x,{k_{\!\perp}}_x,\lambda,a\right)$,
{\it i.e.} by the three momenta, the helicity, and glue, respectively. 
The operator structure of the boson fields is determined by
\begin {equation} 
   \left [a_q,a^\dagger_{q'}\right]  = 
   \delta (k^+-k^{+\,\prime}) 
   \delta ^{(2)}(\vec k_{\!\bot}-\vec k_{\!\bot} ^{\,\prime}) 
   \delta _\lambda ^{\lambda ^\prime} 
   \delta _a^{a^\prime} 
.\end {equation}
Now the space integrations can be performed straight forwardly.
At the end one remains with the $P^\nu$ as operators in Fock-space. 
 
\section{Energy-momentum as Fock-space operators}
 
The three space-like momentum operators $P^+ $ and $\vec P_{\!\perp}$
are diagonal,
\begin{eqnarray}
     P^+ &=& 
     \sum_{\lambda}\int dk^+ d^2{k_{\perp}}\ \left( 
      k^+_q\,b_q^\dagger b_q +
      k^+_q\,d_q^\dagger d_q + 
      k^+_q\,a_q^\dagger a_q \right)
,\\
     \vec P _{\!\perp} &=& 
     \sum_{\lambda}\int dk^+ d^2{k_{\perp}}\ \left( 
     (\vec k_{\!\perp})_q\,b_q^\dagger b_q +
     (\vec k_{\!\perp})_q\,d_q^\dagger d_q + 
     (\vec k_{\!\perp})_q\,a_q^\dagger a_q \right)
.\end{eqnarray}
Here as well as below the summation over color, glue and flavor
is suppressed in the notation. Their eigenvalue
is the same for all Fock states, {\it i.e.}
\begin{equation}
     P^+ = \sum_{i\in\nu}\left(k^+\right)_i,\qquad
     \vec P _{\!\perp} = \sum_{i\in\nu}\left(\vec k_{\!\perp}\right)_i
,\end{equation}
where $i$ runs over all particles in a Fock state $\nu$.
Since $k^+$ and thus $P^+$ are positive, one can introduce 
{\em longitudinal momentum fractions} 
$\displaystyle x_i={k^+_i}/{P^+}$, and, due to boost-invariance, 
{\em intrinsic transversal momenta} ${\vec k_{\!\perp_i}}$.
They obey 
\begin{equation}
     \sum_{i\in\nu}\left(x\right)_i = 1, \qquad
     \sum_{i\in\nu}\left(\vec k_{\!\perp}\right)_i = 0
,\end{equation}
and will be used extensively below. Occasionally, the more compact notation 
\begin{equation}
   \int[d^3q] \equiv \int dx_q \int d^2{k_{\perp}}_q
   \sum_{\lambda_q} \sum_{f_q} \sum_{c_q}
\end{equation}
will be used.

%%%%%%%%%%%%%%%%%%%%%%%%%%%%%%%%%%%%%%%%%%%%%%%%%%%%%%%%%%%%%%%%%%%
%vertex with spinors
\begin{table} 
\caption [verspi] {\label {tab:verspi} 
   The vertex interaction in terms of Dirac spinors.
   The matrix elements $V_{n}$ are displayed on the right, 
   the corresponding (energy) graphs on the left.
   Abbreviation: \\
   $\displaystyle
   \Delta_V = P^+ 
   \frac{g}{\sqrt{16\pi^3}} 
   \delta (k^+_1 - k^+_2 - k^+_3) 
   \delta ^{(2)} (\vec k _{\!\perp,1} - \vec k _{\!\perp,2} -
   \vec k _{\!\perp,3} ) $. 
   \par\vskip 1em}
\begin{center}
\begin{tabular}{|ll|} 
\hline
\begin{tabular}{l}  
\includegraphics{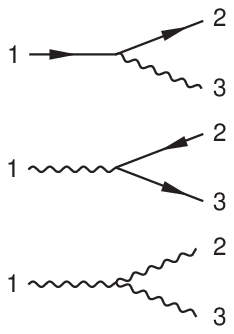}
\\ 
\end{tabular}
&
\begin{tabular}{l}                
  $\displaystyle  V_{1\phantom{,3}} +  
  {\Delta_V\over \sqrt{k^+_1k^+_2 k^+_3}}
  \ (\bar u _1 /\!\!\!\epsilon_3T^{a_3}u_2)$
\\ 
  $\displaystyle V_{3\phantom{,3}} +  
  {\Delta_V \over \sqrt{k^+_1k^+_2 k^+_3}}
  \ (\bar v_2 /\!\!\!\epsilon^\star_1T^{a_1}u_3)$
\\ 
  $\displaystyle V_4 =  
  {iC_{a_2a_3}^{a_1}\,\Delta_V \over \sqrt{k^+_1k^+_2 k^+_3}}
  \ (\epsilon^\star_1k_3)\ (\epsilon_2\epsilon_3) $
\\ 
  $\displaystyle \phantom{V_4} +
  {iC_{a_2a_3}^{a_1}\,\Delta_V \over \sqrt{k^+_1k^+_2 k^+_3}}
  \ (\epsilon_3k_1)\ (\epsilon^\star_1\epsilon_2) $
\\ 
  $\displaystyle \phantom{V_4} +
  {iC_{a_2a_3}^{a_1}\,\Delta_V \over \sqrt{k^+_1k^+_2 k^+_3}}
  \ (\epsilon_3k_2)\ (\epsilon^\star_1\epsilon_2) $
\\ 
\end{tabular}
\\ \hline
\end{tabular}
\end{center}
\end{table}
%%%%%%%%%%%%%%%%%%%%%%%%%%%%%%%%%%%%%%%%%%%%%%%%%%%%%%%%%%%%%%%
The time-like component of $P^\nu$ is the analogue 
to the energy or the Hamiltonian in the 
conventional instant form of Hamiltonian dynamics and is highly off-diagonal.
Rather than $P^-$, however, one considers the operator
of invariant mass squared,
\begin{equation}
     P_\nu P^\nu = P^+P^- - \vec P_{\!\bot} ^2 = P^+P^- \equiv H_{\rm LC}
,\end{equation}
referred to as the `light-cone Hamiltonian $H_{\rm LC}$ \cite{BroPauPin98}.
This is reasonable, since $P^-$ is multiplied only with the 
c-number $P^+$.
Unlike in the instant form, the front form Hamiltonian  
is additive in the free part $T$ and the interaction $U$,
\begin{equation}
     H_{\rm LC} = T + U
.\end{equation}
\textbf{The kinetic energy} $T$ is defined as that part of $H_{\rm LC}$ 
which is independent of the coupling constant and which
can be interpreted as the free invariant mass-squared
of the system. It is the sum of the three 
diagonal operators 
\begin{equation}
     T = \int[d^3q] \ \left( 
      \bigg( {m^2 + \vec k_{\!\bot} ^2 \over x} \bigg) _q b_q^\dagger b_q +
      \bigg( {m^2 + \vec k_{\!\bot} ^2 \over x} \bigg) _q d_q^\dagger d_q + 
      \bigg({\vec k_{\!\bot} ^2 \over x}  \bigg) _q a_q^\dagger a_q \right)
.\label{eq:4kinetic}\end{equation}
%\\
\noindent\textbf{The interaction  energy} $U$ breaks up into 20 different
operators, grouped  into 
\begin{equation}
      U= V + F + S 
.\end{equation}
\textbf{The vertex interaction} $V$ is a sum of  4 operators
\begin{eqnarray}
   V  &=& \int[d^3q_1] \int[d^3q_2] \int[d^3q_3]    
\nonumber\\ &\bigg(& 
   \left[b^\dagger_1b_2a_3\,V_{1}(1;2,3)+{\rm h.c.}\right]+ 
   \left[d^\dagger_1d_2a_3\,V_{2}(1;2,3)+{\rm h.c.}\right] 
\nonumber \\ &+& 
   \left[a^\dagger_1d_2b_3\,V_{3}(1;2,3)+{\rm h.c.}\right]+ 
   \left[a^\dagger_1a_2a_3\,V_{4}(1;2,3)+{\rm h.c.}\right] \bigg)
.\label{eq:vertex_interaction}\end{eqnarray}
It changes the particle number by 1.
The {\em matrix elements} $V_{n}(1;2,3)$ are complex $c$-numbers 
with $V_{2}(1;2,3)=-V_{1}^{\,\star}(1;2,3)  $.
They are functions of the various single-particle momenta 
$k^+,\vec k_{\!\perp}$, helicities, colors and flavors
and are tabulated in Table~\ref{tab:verspi}. 
As compact notation 
$ b_i= {b _q}_i$ and $V_n(1;2,3)=V_n(q_1;q_2,q_3) $ is used.
It should be emphasized that the graphs in these tables
are {\em energy graphs} but {\em not Feynman diagrams}. 
They symbolize {\em matrix elements} but 
{\em not scattering amplitudes}. 
They conserve three-momentum but {\em not four-momentum}.
\\ %\noindent
\textbf{The fork interaction} $F$ is a sum of 6 operators,
%%%%%%%%%%%%%%%%%%%%%%%%%%%%%%%%%%%%%%%%%%%%%%%%%%%%%%%%%%%%%%%%%%%
% spinor fork [t]
\begin{table}
\caption [forspi] {\label {tab:forspi} 
   The fork interaction in terms of Dirac spinors.
   The matrix elements $F_{n,j}$ are displayed on the right, 
   the corresponding (energy) graphs on the left.
   Abbreviation: \\
   $\displaystyle
   \Delta = 
   \frac{P^+}{2}\,\frac{g^2}{16\pi^3}
   \delta (k^+_1 + k^+_2 - k^+_3-k^+_4) 
   \delta ^{(2)} (\vec k _{\!\perp,1} + \vec k _{\!\perp,2} -
                  \vec k _{\!\perp,3} - \vec k _{\!\perp,4} ) $. 
   \par\vskip 1em}
\begin{center}
\begin{tabular}{|ll|} 
\hline 
\begin{tabular}{l}  
\includegraphics{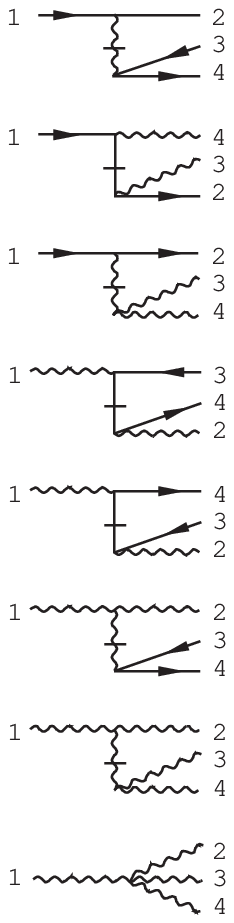}
\end{tabular}
&
\begin{tabular}{l}  
  $\displaystyle F_{1\phantom{,1}} = + 
  {2\Delta\over \sqrt{k^+_1k^+_2 k^+_3 k^+_4}}
  {(\bar u _1T^a\gamma^+u_2)\ (\bar v_3\gamma^+T^au_4)
   \over(k^+_1-k^+_2)^2}$
\\                                           
  $\displaystyle F_{3,1} = +
  {\Delta\over \sqrt{k^+_1k^+_2 k^+_3 k^+_4}}
  {(\bar u _1T^{a_4}/\!\!\!\epsilon_4\gamma^+
   /\!\!\!\epsilon_3T^{a_2}u_2)
   \over( k^+_1-k^+_4)}$
\\                                           
  $\displaystyle F_{3,2} = -
  {2k^+_3\Delta\over \sqrt{k^+_1k^+_2 k^+_3 k^+_4}}
  {(\bar u _1T^a\gamma^+u_2)
  \ (\epsilon_3iC^a\epsilon_4)\over( k^+_1-k^+_2)^2}$
\\                                           
  $\displaystyle F_{5,1} = +
  {\Delta\over \sqrt{k^+_1k^+_2 k^+_3 k^+_4}}
  {(\bar v _3T^{a_1}/\!\!\!\epsilon^\star_1\gamma^+
  /\!\!\!\epsilon_2T^{a_2}u_4)
  \over( k^+_1-k^+_3)}$
\\                                           
  $\displaystyle F_{5,2} = -
  {\Delta\over \sqrt{k^+_1k^+_2 k^+_3 k^+_4}}
  {(\bar v _3T^{a_2}/\!\!\!\epsilon_2\gamma^+
  /\!\!\!\epsilon^\star_1T^{a_1}u_4)
  \over( k^+_1-k^+_4)}$
\\                                           
  $\displaystyle F_{5,3} = +
  {2(k^+_1+k^+_2)\Delta\over \sqrt{k^+_1k^+_2 k^+_3 k^+_4}}
  {(\bar v _3T^a\,\gamma^+\,u_4)   
   \ (\epsilon^\star_1iC^a\epsilon_2)  \over( k^+_1-k^+_2)^2}$
\\                                           
  $\displaystyle F_{6,1} = +
  {2k^+_3(k^+_1+k^+_2)\Delta\over 
  \sqrt{k^+_1k^+_2 k^+_3 k^+_4}\hfill}
  \ {(\epsilon^\star_1C^a\epsilon_2)
  \   (\epsilon_3C^a\epsilon_4)  \over( k^+_1-k^+_2)^2}$
\\                                           
  $\displaystyle F_{6,2} = +
  {2\Delta\over \sqrt{k^+_1k^+_2 k^+_3 k^+_4}}
  \ (\epsilon^\star_1\epsilon_3)\ (\epsilon_2\epsilon_4) 
  \ C^a_{a_1a_2} C^a_{a_3a_4}$
\end{tabular}
\\  \hline
\end{tabular}
\end{center}
\end{table}
%%%%%%%%%%%%%%%%%%%%%%%%%%%%%%%%%%%%%%%%%%%%%%%%%%%%%%%%%%%%%%%%%%%
\begin{eqnarray}
   F &=& \int[d^3q_1] \int[d^3q_2] \int[d^3q_3] \int[d^3q_4] 
\nonumber \\ &\bigg(&
   \left[b_1 ^\dagger b_2 d_3 b_4\,F_1(1;2,3,4)+{\rm h.c.}\right]+ 
   \left[d_1 ^\dagger d_2 b_3 d_4\,F_2(1;2,3,4)+{\rm h.c.}\right]   
\nonumber \\ &+&
   \left[b_1 ^\dagger b_2 a_3 a_4\,F_3(1;2,3,4)+{\rm h.c.}\right]+
   \left[d_1 ^\dagger d_2 a_3 a_4\,F_4(1;2,3,4)+{\rm h.c.}\right]  
\nonumber \\ &+&
   \left[a_1 ^\dagger a_2 d_3 b_4\,F_5(1;2,3,4)+{\rm h.c.}\right]+
   \left[a_1 ^\dagger a_2 a_3 a_4\,F_6(1;2,3,4)+{\rm h.c.}\right] \bigg) 
.\end{eqnarray}
It changes the particle number  by 2. 
The matrix elements $F_n(1;2,3,4)$ and their graphs are 
tabulated in Table~\ref{tab:forspi},
with $F_2=F_1 $ and $F_4=F_3 $.
%%%%%%%%%%%%%%%%%%%%%%%%%%%%%%%%%%%%%%%%%%%%%%%%%%%%%%%%%%%%%%%%%%%
%spinor seagulls
\begin{table}
\caption [seaspi] {\label {tab:seaspi} 
   The seagull interaction in terms of Dirac spinors.
   The matrix elements $S_{n,j}$ are displayed on the right, 
   the corresponding (energy) graphs on the left.
   Abbreviation: \\
   $\displaystyle
   \Delta = 
   \frac{P^+}{2}\,\frac{g^2}{16\pi^3}
   \delta (k^+_1 + k^+_2 - k^+_3-k^+_4) 
   \delta ^{(2)} (\vec k _{\!\perp,1} + \vec k _{\!\perp,2} -
                  \vec k _{\!\perp,3} - \vec k _{\!\perp,4} ) $. 
   \par\vskip 1em}
\begin{center}
\begin{tabular}{|ll|} 
\hline 
\begin{tabular}{l}  
\includegraphics{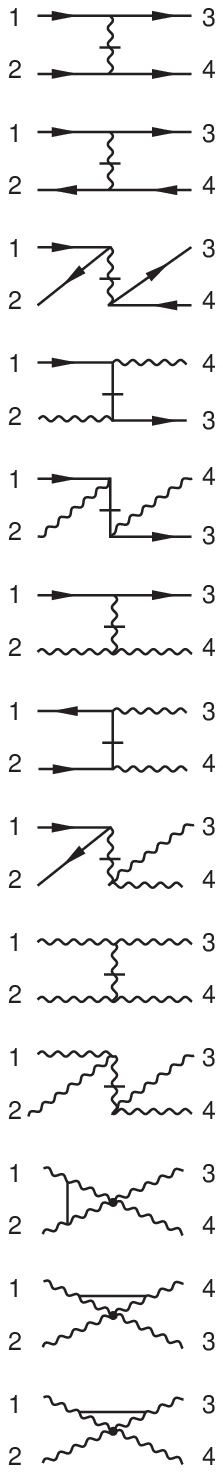}
\\ 
\end{tabular}
&
\begin{tabular}{l}  
  $\displaystyle S_{1\phantom{,1}} = - 
  {\Delta\over \sqrt{k^+_1k^+_2 k^+_3 k^+_4}}
  {(\bar u _1T^a\gamma^+u_3)\ (\bar u _2\gamma^+T^au_4)
   \over(k^+_1-k^+_3)^2}$
\\ 
  $\displaystyle S_{3,1} = +
  {2\Delta\over \sqrt{k^+_1k^+_2 k^+_3 k^+_4}}
  {(\bar u _1T^a\gamma^+u_3)\ (\bar v _2\gamma^+T^av_4)
  \over(k^+_1-k^+_3)^2}$
\\ 
  $\displaystyle S_{3,2} = -
  {2\Delta\over \sqrt{k^+_1k^+_2 k^+_3 k^+_4}}
  {(\bar v_2T^a\gamma^+u_1)\ (\bar v _4\gamma^+T^au_3)
  \over(k^+_1+k^+_2)^2}$
\\ 
  $\displaystyle S_{4,1} = +
  {\Delta\over \sqrt{k^+_1k^+_2 k^+_3 k^+_4}}
  {(\bar u _1T^{a_4}/\!\!\!\epsilon_4\gamma^+
   /\!\!\!\epsilon^\star _2T^{a_2}u_3)
   \over( k^+_1-k^+_4)}$
\\ 
  $\displaystyle S_{4,2} = +
  {\Delta\over \sqrt{k^+_1k^+_2 k^+_3 k^+_4}}
  {(\bar u _1T^{a_2}/\!\!\!\epsilon^\star _2\gamma^+
  /\!\!\!\epsilon_4T^{a_4}u_3)
  \over( k^+_1+k^+_2)}$
\\ 
  $\displaystyle S_{4,3} = +
  {2(k^+_2+k^+_4)\Delta\over \sqrt{k^+_1k^+_2 k^+_3 k^+_4}}
  {(\bar u _1T^a\gamma^+u_3)
  \ (\epsilon_2^\star iC^a\epsilon_4)\over( k^+_1-k^+_3)^2}$
\\ 
  $\displaystyle S_{6,1} = +
  {\Delta\over \sqrt{k^+_1k^+_2 k^+_3 k^+_4}}
  {(\bar u _1T^{a_3}/\!\!\!\epsilon_3\gamma^+/\!\!\!\epsilon_4
   T^{a_4}v_2) \over( k^+_1-k^+_3)}$
\\ 
  $\displaystyle S_{6,2} = -
  {(k^+_3-k^+_4)\Delta\over \sqrt{k^+_1k^+_2 k^+_3 k^+_4}}
  {(\bar u _1T^a\,\gamma^+\,v_2)   
   \ (\epsilon_3iC^a\epsilon_4)  \over( k^+_1+k^+_2)^2}$
\\ 
  $\displaystyle S_{7,1} = -
  {(k^+_1+k^+_3)(k^+_2+k^+_4)\Delta\over 
  \sqrt{k^+_1k^+_2 k^+_3 k^+_4}\hfill}
  \ {(\epsilon^\star_1C^a\epsilon_3)
   \ (\epsilon^\star_2C^a\epsilon_4)  \over( k^+_1-k^+_3)^2}$
\\ 
  $\displaystyle S_{7,2} = +
  {2k^+_3k^+_4\Delta\over \sqrt{k^+_1k^+_2 k^+_3 k^+_4}}
   \ {(\epsilon^\star_1C^a\epsilon^\star_2)
    \ (\epsilon_3C^a\epsilon_4)  \over( k^+_1+k^+_2)^2}$
\\ 
  $\displaystyle S_{7,3} = +
  {\Delta\over \sqrt{k^+_1k^+_2 k^+_3 k^+_4}}
  \ (\epsilon^\star_1\epsilon_3)\ (\epsilon^\star_2\epsilon_4) 
  \ C^a_{a_1a_2} C^a_{a_3a_4}$
\\ 
  $\displaystyle S_{7,4} = +
  {\Delta\over \sqrt{k^+_1k^+_2 k^+_3 k^+_4}}
  \ (\epsilon^\star_1\epsilon_3)\ (\epsilon^\star_2\epsilon_4) 
  \ C^a_{a_1a_4} C^a_{a_3a_2}$
\\ 
  $\displaystyle S_{7,5} = +
  {\Delta\over \sqrt{k^+_1k^+_2 k^+_3 k^+_4}}
  \ (\epsilon^\star_1\epsilon^\star_2)\ (\epsilon_3\epsilon_4) 
  \ C^a_{a_1a_3} C^a_{a_2a_4}$
\\ 
\end{tabular}
\\  \hline
\end{tabular}
\end{center}
\end{table}
%
%%%%%%%%%%%%%%%%%%%%%%%%%%%%%%%%%%%%%%%%%%%%%%%%%%%%%%%%%%%%%%%%%%%
\\ %\noindent
\textbf{The seagull interaction} $S$ is a sum of 7 operators  
\begin{eqnarray}
      S  &=& \int[d^3q_1] \int[d^3q_2] \int[d^3q_3] \int[d^3q_4] 
\nonumber \\ &\bigg(&
      b_1^\dagger b_2^\dagger b_3 b_4\,S_1(1,2;3,4) + 
      d_1^\dagger d_2^\dagger d_3 d_4\,S_2(1,2;3,4) +
      b_1^\dagger d_2^\dagger b_3 d_4\,S_3(1,2;3,4)  
\nonumber \\ &+&
      b_1^\dagger a_2^\dagger b_3 a_4\,S_4(1,2;3,4) +
      d_1^\dagger a_2^\dagger d_3 a_4\,S_5(1,2;3,4) +
\nonumber \\ &+&
     [b_1^\dagger d_2^\dagger a_3 a_4\,S_6(1,2;3,4) +{\rm h.c.}] +
      a_1^\dagger a_2^\dagger a_3 a_4\,S_7(1,2;3,4) \bigg)
.\end{eqnarray}
It does not change particle number.
The matrix elements $S_{n}(1,2;3,4) $ and their energy graphs
are tabulated in Table~\ref{tab:seaspi},
with $S_2=S_1$ and $S_5=S_4$.
%   In all of these tables the some conventions are used,
%   for instance$F_5=F_{5,1}+F_{5,2}+F_{5,3}$.
\\
\noindent\textbf{Summarizing} these considerations, one can state that 
the light-cone Hamiltonian $H\equiv H_{\rm LC}$ consists
of 23 operators with different operator structure. 

\section{The Sawicki transformation}
\label{sec:saw}

When dealing with practical matters in the front form, 
particularly when working numerically, one is often pondered 
by the fact that the kinematical variables have 
a different range: 
$0 \leq x \leq 1$ and $-\infty \leq \vec k_{\!\perp} \leq \infty$. 
An elegant escape is to use a variable transform 
from $x$ to $k_z$ whose precursor had been introduced first by
Sawicki \cite{Saw85}.

The Sawicki transformation is demonstrated here at hand of 
the example: 
\begin{eqnarray} 
    M^2\psi(x,\vec{k}_{\!\perp})  
    = \left[ 
    \frac{m^2_{1} + \vec k_{\!\perp}^{\,2}}{x} +
    \frac{m^2_{2} + \vec k_{\!\perp}^{\,2}}{1-x}  
    \right] &\psi(x,\vec{k}_{\!\perp})&   
\nonumber\\ 
     - \frac{m_1 m_2}{\pi^2} 
    \int\limits_{0}^{1} dx' 
    \int\limits_{-\infty}^{\infty} d^2 \vec k_{\!\perp}^\prime
    \ \frac{1}{\sqrt{ x(1-x) x^\prime(1-x^\prime)}}
    \ \frac{\alpha}{Q^2} 
    \ &\psi(x',\vec{k}_{\!\perp}')&
.\label{eq:1}\end{eqnarray} 
The physical interpretation of this integral equation in the three
variables $x$ and $\vec{k}_{\!\perp}$ is given below.
$M ^2$ is the invariant-mass squared eigenvalue and 
$\psi(x,\vec{k}_{\!\perp})$ is the associated eigenfunction. 
It is the probability amplitude for
finding a particle of mass $m _1$ with momentum fraction $x$ 
and transversal momentum $\vec k_{\!\perp}$,
and correspondingly an anti-particle of mass $m _2$ 
with $1-x$ and $-\vec k_{\!\perp}$.
The inverse of the kernel is defined as 
\begin{equation}
   Q ^2 (x,\vec k_{\!\perp},x',\vec k_{\!\perp}') 
   = -\frac{1}{2}\left[(k_{1}-k_{1}')^2 + (k_{2}-k_{2}')^2\right]
,\end{equation}
thus
\begin{eqnarray}
   Q ^2 = \left(\vec k_{\!\perp}-\vec k_{\!\perp}'\right)^2 
   &-& \frac{1}{2}\left(x-x'\right)\left(
   \frac{m_1^2+{k_{\!\perp}}^2}{x }-
   \frac{m_1^2+{k_{\!\perp}'}^2}{x'}\right) 
\nonumber\\
   &+& \frac{1}{2}\left(x-x'\right)\left(
   \frac{m_2^2+{k_{\!\perp}}^2}{1-x }-
   \frac{m_2^2+{k_{\!\perp}'}^2}{1-x'}\right)
.\end{eqnarray} 
The coupling constant is $\alpha$. 

Integration variables can be changed from $x$ to $k_z$ by introducing 
\begin{equation}
    x(k_z)  = \frac{E_1+k_z}{E_1+E_2},\mbox{ with }
    E_{1,2} = \sqrt{m^{\,2}+ k_z^2 + \vec k_{\!\perp}^{\,2}}
.\end{equation}
The new integration variable $k_z$ can be combined with 
$\vec k_{\!\perp}$ into $\vec k= (\vec k_{\!\perp},k_z)$
which however is not a 3-vector in the usual sense.
The Jacobian is 
\begin{equation}
    \frac {dx}{x(1-x)} = \frac {1} {A ({k})}\frac {dk_z}{m_r} 
,\end{equation}
with the dimensionless function 
\begin{equation}
    A (k) = \frac{1}{m_r} \frac{E_1E_2}{E_1+E_2} 
.\end{equation}
The reduced mass and the sum mass is 
\begin{equation}
    \frac{1}{m_r} = \frac{1}{m_1} + \frac{1}{m_2}
    ,\hskip4em
    m_s =  m_1 + m_2
.\end{equation}
The free invariant mass-squared
on the r.h.s. of Eq.(\ref{eq:1}), {\it i.e.}
\begin{equation}
    \frac{m^2_{1} + \vec k_{\!\perp}^{\,2}}{x} +
    \frac{m^2_{2} + \vec k_{\!\perp}^{\,2}}{1-x}
,\end{equation} 
plays the role of a the kinetic energy $T$. 
Rewriting it, one uses first
\begin{equation}
    \frac{m^2_{1} + \vec k_{\!\perp}^{\,2}} {x} =
    \frac{m^2_{1} + \vec k_{\!\perp}^{\,2}} {E_1+k_z} \left(E_1+E_2\right) =
    \left(E_1-k_z\right) \left(E_1+E_2\right) 
,\end{equation} 
and correspondingly
\begin{equation}
    \frac{m^2_{1} + \vec k_{\!\perp}^{\,2}} {1-x} =
    \left(E_2+k_z\right) \left(E_1+E_2\right) 
.\end{equation} 
The kinetic energy can thus be rewritten identically as
\begin{equation}
    T (k) =  
    \frac{ m^2_{1} + \vec k_{\!\perp}^{\,2}}{x} +
    \frac{ m^2_{2} + \vec k_{\!\perp}^{\,2}}{1-x} - m_s^2 
    \equiv  C(k)\,\vec k ^2
,\end{equation} 
with the dimensionless function
\begin{equation}
    C(k) = (E_{1}+ m_{1} + E_{2}+  m_{2})  
    \left(
    \frac{1}{E_{1}+ m_{1}} +
    \frac{1}{E_{2}+ m_{2}}\right) 
.\end{equation}
If the wave function is substituted
according to 
\begin{equation}
   \psi(x,\vec k _{\!\perp}) =
   \phi(x,\vec k _{\!\perp})
   \ \sqrt{\frac{A(x,\vec{k}_{\!\perp})}{x(1-x)}} 
,\label{eq:phi}\end{equation}
one converts the integral equation (\ref{eq:1}) into an integral equation 
\begin{equation}
   \left[M^2 - m_s^2 - C(k)\vec k ^2 \right]\phi(\vec k) 
   = - \frac{m_s}{\pi^2} \int
   \frac {d^3\vec k'} {\sqrt{A(k)A(k')}} 
   \ \frac{\alpha}{Q^2(\vec k,\vec k')} 
   \ \phi(\vec k') 
,\label{eq:12}\end{equation}
like in usual momentum space. 
However, since no changes have been made except relabeling the
integration variable, it is identical with Eq.(\ref{eq:1}), 
and thus is a genuine front-form or light-cone integral equation.

The above simplifies considerably for equal particle masses 
$m_1=m_2=m$:
\begin{equation}
   m_r = \frac{m}{2},\qquad m_s = 2m
,\end{equation}
and
\begin{equation}
   x(k_z) = \frac{1}{2}\left(1-
   \frac{k_z}{\sqrt{m^2 + \vec k_{\!\perp}^{\,2} + k_z^2}}\right)
   ,\quad
   k_z(x) = \left(x-\frac{1}{2}\right)
   \sqrt{\frac{m^2+k _{\!\perp}^2}{x(1-x)}}
.\label{eq:c82}\end{equation}
The coefficient functions $A$ and $C$ become 
\begin{equation}
   A(k) = \sqrt{1+\frac{k^2}{m^2}},\qquad
   C(k) = 4
.\end{equation}
In a non-relativistic situation holds $k^2\ll m^2$, thus 
\begin{equation}
   A(k)\sim 1,\qquad
   Q^2\sim(\vec k-\vec k')^2
.\end{equation}
To substitute that in the kernel of an integral equation 
like (\ref{eq:12})
is certainly not justified, since the integration variable 
has $k'\rightarrow\infty$ at the upper limit. 
But if one does it anyway, {\it i.e.}
\begin{equation}
   \left[M^2 - 4m^2 - 4\vec k ^2 \right]\phi(\vec k) 
   = - \frac{2m}{\pi^2} \int d^3\vec k'
   \ \frac{\alpha}{\left(\vec k-\vec k'\right)^2} 
   \ \phi(\vec k') 
,\label{eq:14}\end{equation}
one can generate analytical solutions, see Sect.~\ref{sec:fourier}.
The ground state has
\begin{equation}
  M^2 = m^2(4-\alpha^2),\qquad 
  \phi(k) = \frac{1}{\left(1+\displaystyle\frac{k^2}{p_B^2}\right)^2}
,\end{equation}
with $p_B=m\alpha/2$.
When substituting the inverse Sawicki transformation back
into the defining Eq.(\ref{eq:phi}), 
\begin{equation}
  \vec k ^2 = \vec k_{\!\perp}^{\,2} + k_z^2 =
  \frac{m^2\left(2x-1\right)^2+\vec k_{\!\perp}^{\,2}}{4x(1-x)}
,\end{equation}
one expresses the (un-normalized) light-cone wavefunction
in terms of the light-cone variables $x$ and $\vec k_{\!\perp}$,
\begin{eqnarray}
   \psi(x,\vec k _{\!\perp}) &=&
   \frac{1}{\sqrt{x(1-x)}} 
   \frac{1}{\left(1+\displaystyle
   \frac{m^2\left(2x-1\right)^2+\vec k_{\!\perp}^{\,2}}
   {\alpha^2 m^2\,x(1-x)}
   \right)^2} 
\nonumber\\ &\times&
   \left(1+\displaystyle
   \frac{m^2\left(2x-1\right)^2+\vec k_{\!\perp}^{\,2}}
   {4 m^2\,x(1-x)}
   \right)^{\frac{1}{4}}
.\end{eqnarray}
The last factor can also be set to unity because of $A\sim 1$.
Even then, the present $\psi(x,\vec k _{\!\perp})$ 
differs considerably from a similar expression
in the literature \cite{leb80}. In particular the factor 
$\sqrt{x(1-x)}$ is different. Here is the part of 
$\psi(x,\vec k _{\!\perp})$ which violates rotational invariance
as it should!

\section{Fourier transforms in the front form}
\label{sec:fourier}

%%%%%%%%%%%%%%%%%%%%%%%%%%%%%%%%%%%%%%%%%%%%%%%%%%%%%%%%%%%%%%%%%%%
%    Fourier transforms 
\begin{table}
\caption [Fourier] {\label {tab:fourier} 
   Compilation of a few familiar Fourier transforms.
   \par\vskip 1em}
\begin{center}
\begin{tabular}{|c|c|c|c|} 
   \hline \hline
   \rule[-3ex]{0ex}{9ex}
   $\displaystyle U(\vec q)$ & 
   $\displaystyle F[U]=\int\!d^3 \vec q\ \mbox{e}^{-i\vec q\vec x}\ U(\vec q)$ &
   $\displaystyle U(\vec q)$ & 
   $\displaystyle F[U]=\int\!d^3 \vec q\ \mbox{e}^{-i\vec q\vec x}\ U(\vec q)$
\\ \hline \hline
   \rule[-3ex]{0ex}{9ex}
   $\displaystyle \frac{1}{q^2 + \mu^2}$ & 
   $\displaystyle \frac{2\pi^2}{r}\ \mbox{e}^{-\mu r}$ &
   $\displaystyle \frac{1}{q^2}$ & 
   $\displaystyle \frac{2\pi^2}{r} $ 
\\ 
   \rule[-3ex]{0ex}{9ex}
   $\displaystyle \frac{\mu^2}{\left(q^2 + \mu^2\right)^2}$ & 
   $\displaystyle \pi^2\mu\ \mbox{e}^{-\mu r} $ &
   $\displaystyle \frac{1}{\mu^2}$ & 
   $\displaystyle \frac{(2\pi)^3}{\mu^2}\ \delta ^{{(3)}} (\vec x)$ 
\\ 
\hline \hline
\end{tabular}
\end{center}
\end{table}
%    
%%%%%%%%%%%%%%%%%%%%%%%%%%%%%%%%%%%%%%%%%%%%%%%%%%%%%%%%%%%%%%%%%%%
The interpretation of front-form equations such as Eq.(\ref{eq:1})
in terms of physics is not always easy,
even if they are Sawicki-transformed to a momentum space integral
equation like (\ref{eq:12}). 
Our intuition and our language is shaped in configuration space.
To boost that one can apply Fourier transformation, 
but only in principle, since an exact but analytical Fourier 
transformation of Eq.(\ref{eq:12}) for example is close to impossible.
But one can apply simplifying assumptions. 

The steps are demonstrated at hand of the example 
\begin{equation}
   \left[M^2-m_s^2-\frac{m_s}{m_r}\vec{k}^2 \right]\phi(\vec{k}) 
   = - \frac{m_s}{\pi^2} \int d^3\vec{k'} 
   \left(\frac{\alpha}{(\vec k - \vec k')^2} + 
   \frac{\alpha}{2m_r m_s}\right) 
   \,\phi(\vec{k'})
.\end{equation}
It is identical with Eq.(\ref{eq:12}), except for the 
simple additive constant $\alpha/(2m_r m_s)$ in the kernel.
What is its interpretation?

Transforming the momentum-space wave function to one 
in configuration space by
$   \psi(\vec x)=(2\pi)^{-\frac{3}{2}} 
   \int\! d^3 \vec p \ \mbox{e}^{-i\vec p\vec x}\ \phi(\vec p)$,
and using the expressions compiled in Table~\ref{tab:fourier}, one gets
\begin{equation}
   \left[M^2-m_s^2 +\frac{m_s}{m_r}\vec \nabla^{\,2}\right]\psi(\vec r) = 
   2m_s V(r) \psi(\vec r)
.\end{equation}
The function $V(r)$ has the dimension of a mass and
is defined by
\begin{equation}
   V (r) = - \frac{\alpha}{r} - 
   2\frac{\alpha\pi}{m_r m_s}\delta^{(3)}(\vec r) 
.\label{eq:16}\end{equation}
The strange `2'  finds its explanation 
as the gyro-magnetic ratio for a fermion, 
by analogy with the hyperfine interaction 
in hydrogen in the singlet channel,
\begin{equation}
   V_{\mbox{hf-s}} (r) =  - \frac{\alpha}{r} -  
   g_p\frac{\alpha\pi}{m_e m_p}\delta^{(3)}(\vec r)
.\end{equation}
Here, $m_p$ and $g_p$ are the protons mass and gyro-magnetic ratio,
respectively and $m_e$ is the  mass of the electron.
If the mass-square eigenvalue is substituted
by an energy eigenvalue $E$, defined by
\begin{equation}
   M^2 = m_s^2 + m_s E
,\end{equation}
one gets the conventional Schr\"odinger equation
for the Coulomb plus a contact potential,
\begin{equation}
   -\frac{\nabla^{\,2}}{2m_r}\vec \psi(\vec r) + V(r)\psi(\vec r)  =
   E \psi(\vec r)
.\end{equation}
If the latter is suppressed,
the analytical solution for the ground state is
\begin{equation}
   \psi(\vec r) = \mbox{e}^{-p_B\vec r}
   ,\hskip4em
   E = \frac{p_B^2}{2m_r}
,\end{equation}
with the Bohr momentum $p_B=m_r\alpha$.
If one includes the contact term, one gets in first order 
perturbation theory the usual
\begin{equation}
    E_{\mbox{perturbative}} = \frac{p_B^2}{2m_r} + 
    \alpha \frac{2\pi }{m_r m_s}|\psi(0)|^2
.\end{equation}
The size in either case is
\begin{equation}
    <r^2> = \frac{3}{p_B^2} = \frac{3}{m_r^2\alpha^2} 
.\label{eq:r}\end{equation}
But Eq.(\ref{eq:16}) has no solution in the proper sense! 
A Dirac-delta function is no proper function and must be regulated,
for instance by a Yukawa potential 
\begin{equation}
   V (r) = - \frac{\alpha}{r} -
   \frac{\alpha\mu^2}{m_r m_s}\, \frac{e^{-\mu r}}{r}
.\label{eq:19}\end{equation}
Transforming back to momentum space gives
\begin{equation}
   \left[M^2-m_s^2-\frac{m_s}{m_r}\vec k^2\right]\phi(\vec{k}) 
   = - 
   \frac{\alpha}{\pi^2}\int\!\! d^3\vec{k'} 
   \left(\frac{m_s}{Q^2} + \frac{1}{2m_r}
   \frac{\mu^2}{\mu^2+Q^2} \right)\phi(\vec{k'})
.\label{eq:20}\end{equation}
Note that the coupling constant of the Yukawa potential
$\beta= \alpha \mu^2/(m_r m_s)$ 
is potentially very large, since one must go to the
limit $\mu \longrightarrow \infty$ to retrieve the 
Dirac-delta-function.
It is not trivial to solve Eqs.(\ref{eq:19}) or (\ref{eq:20}).

\end{document}